\documentclass{ws-procs9x6}

\begin{document}
\def\Journal#1#2#3#4{{#1} {\bf #2}, #3 (#4)}
% Some useful journal names
\def\NCA{\em Nuovo Cimento}
\def\NIM{\em Nucl. Instrum. Methods}
\def\NIMA{{\em Nucl. Instrum. Methods} A}
\def\NPA{{\em Nucl. Phys.} A}
\def\NPB{{\em Nucl. Phys.} B}
\def\PLB{{\em Phys. Lett.}  B}
\def\PRL{\em Phys. Rev. Lett.}
\def\PRD{{\em Phys. Rev.} D}
\def\ZPC{{\em Z. Phys.} C}

\title{ %TESTING SUGRA AND STRINGS
TESTING SUGRA UNIFIED MODELS 
\footnote{\uppercase{C}ontribution to Proceedings of the
\uppercase{N}ath\uppercase{F}est in \uppercase{PASCOS'04}}}

\author{UTPAL CHATTOPADHYAY}

\address{Department of Theoretical Physics \\
Indian Association for the Cultivation of Science \\ 
Jadavpur, Kolkata 700032, INDIA\\ 
E-mail: tpuc@iacs.res.in}

\maketitle

\abstracts{
A brief review is given of the ways of testing SUGRA unified models and 
a class of string models using 
data from precision electroweak experiments, Yukawa unification constraints,
and constraints from dark matter experiments. Models discussed in 
detail include mSUGRA, 
extended SUGRA model with non-universalities within $SO(10)$ grand unification, and 
effective theories with modular invariant soft breaking within a generic 
heterotic string framework.
 The implications of the Hyperbolic Branch including the focus point and inversion regions
 for the discovery of supersymmetry in collider experiments and for the 
detection of dark matter in the direct detection experiments are also discussed.
}

\section{Introduction}
\begin{center}
{ It is a pleasure to dedicate this paper as a tribute to my thesis advisor 
Pran Nath. }
\end{center}

Over the past three decades supersymmetry\cite{Wess:1992cp} has come to play a central role in our
exploration of  physics beyond the standard model. However, a 
phenomenologically
viable breaking of supersymmetry requires the framework of local supersymmetry 
which leads to supergravity.   
%i.e., supergravity and string theory.  
Considerable progress has taken place over 
the recent past in building  models within (applied) supergravity 
and in string theory. In this paper 
we give here a brief analysis of testing  supergravity unified 
models\cite{Chamseddine:1982jx}.  We will also extend our discussion to low energy 
signatures 
of a class of models based on heterotic strings.  
%string models.  
First we will focus on supergravity (SUGRA) models, specifically the minimal 
supergravity (mSUGRA) and 
its extensions, which are currently among the leading candidates for 
physics beyond the standard model.  A remarkable 
aspect of mSUGRA model is that the number of
arbitrary parameters that appear in soft breaking in this model 
 is much smaller\cite{giradello:1982} than what is allowed in the 
minimal supersymmetric standard model (MSSM).
Further, within mSUGRA one can achieve a unification 
of the gauge coupling constants\cite{Dienes:1996du} and any small discrepancy between theory and experiment
can also be resolved through gravitational corrections\cite{hall:unification}.
We begin by recalling briefly the basic ingredients
that enter in the construction of supergravity unified models. The construction utilizes applied
supergravity techniques\cite{Chamseddine:1982jx,appliedPN,coupsug}
 coupling an arbitrary set of chiral superfields with a vector multiplet
in the adjoint representation of the gauge group 
and then coupling the system with supergravity.
As is well known the resulting Lagrangian depends on just three arbitrary functions: 
the gauge kinetic energy function $f_{\alpha\beta}$, the K\"ahler potential $K(z_i, z_i^{\dagger})$
and the superpotential $W(z_i)$. Although there are a variety of scenarios such as 
gravity mediation, gauge mediation, anomaly mediation, breaking by anomalous $U(1)$'s etc, 
for the breaking of supersymmetry we focus here on the gravity mediated breaking. This
utilizes a hidden sector and a  visible sector, where supersymmetry is broken in the 
hidden sector and communicated to the visible sector by 
gravity\cite{Chamseddine:1982jx,Barbieri:1982eh}.  In the minimal version
of the model based on a flat K\"ahler potential and a flat gauge kinetic energy function,
the soft breaking sector of the theory is parametrized by the following 
parameters\cite{Nath:1983aw,Hall:1983iz}:
the universal scalar mass $m_0$, the universal gaugino mass $m_{1/2}$, the universal trilinear
coupling $A_0$, the bilinear coupling $B_0$,  and the Higgs mixing parameter 
 $\mu_0$ where $\mu_0$ enters in the superpotential in the form $\mu_0 H_1H_2$.  
Here $H_2$ gives
 mass to the up quark and $H_1$ gives mass to the down quark and the lepton. 

One of the remarkable aspects of  SUGRA model observed early on was the phenomenon that
soft breaking can trigger breaking of the electroweak symmetry\cite{Chamseddine:1982jx}.  
This phenomenon appears
quite naturally when the renormalization group effects are included as one moves from the 
grand unification scale to the electroweak 
scale\cite{grandunisug,inoue,oldrge,newrge}.  
In this case the Higgs doublet mass-square, 
$m_{H_2}^2$  which turns negative triggers a breaking of the electroweak
symmetry. In the analysis one takes into account also the one loop corrections to the 
effective potential\cite{effpot1,effpot2}. 
The minimization of the potential gives rise to two constraints 
arising from the conditions $\partial_{<H_1^0>}V=0= \partial_{<H_2^0>}V$.  
One of these 
constraints can be used to determine $|\mu_0|$ while the other 
allows one to eliminate $B_0$ 
in terms of $\tan\beta=<H_2>/<H_1>$.
Thus after electroweak symmetry breaking 
mSUGRA is described by the parameters: $m_0$, $m_{1/2}$, $A_0$, $\tan\beta$, and sign($\mu$).
In extended supergravity models one can include non-universalities in the Higgs sector and 
non-universalities of the gaugino masses.  
Additionally there is also the possibility of 
extending SUGRA models to include CP phases.  
Large CP phases can indeed be compatible with the electric dipole moment
(EDM) constraints\cite{tipnedm}.  In mSUGRA there are two such phases, but
more phases appear in extended SUGRA models. These phases can affect a variety of low
energy phenomenon such as analyses of dark matter. 
Another class of models studied widely are a variety of string based models, 
specifically
models based on the heterotic string\cite{heterotic}. Of special interest to us here is the low energy 
effective theory that results from these models. 

Such a low energy effective
theory should take into account as much of the symmetry of the underlying string model
as possible.  One of these is the  $SL(2,Z)$ large radius - small radius duality symmetry.
Such a symmetry may be valid even non-perturbatively and thus it is reasonable to probe 
effective low energy supergravity theories which possess modular invariance. 
% The analyses in  SUGRA and string models must of course be constrained by 
Investigation must of course include the stringent 
 experimental constraints such as the flavor changing neutral current constraint given by
 the branching ratio 
$Br(b\rightarrow s+\gamma)$\cite{bsgamma1,bsgamma2,bsgamma3,bsgammaExpt}.  
The allowed parameter space of such models
 must also be consistent with the current constraints from the muon 
anomalous magnetic moment. 
 Further, with R-parity invariance the lightest supersymmetric particle 
(LSP) in such models is absolutely stable. In SUGRA models
 and also in a class of string models the LSP in a significant part of the parameter space  
 is the lightest neutralino, which is a strong candidate for cold dark matter (CDM). 
%Thus the lightest neutralino turns out to be a candidate for cold
% dark matter in much of the parameter space of SUGRA models and also in a class of string models.
 
The rest of the paper is organized as follows.  
In section Sec.2 we discuss the radiative 
 electroweak symmetry breaking and discuss the hyperbolic 
 branch of REWSB in SUGRA type of models 
in this context. In Sec.3 we discuss the precision data
 constraints.  In Sec.4 we discuss the constraints of 
 Yukawa unification on SUGRA models.  
In Sec.5 we discuss the constraints of dark matter 
 and the possibility of detection of dark matter predicted by SUGRA and string models 
 in direct detection experiments. 
 Conclusions are given in Sec.6
           
%\section{HYPERBOLIC BRANCH IN SUGRA AND STRING MODELS} 
\section{Hyperbolic Branch from Radiative Electroweak Symmetry Breaking} 
\label{HBFPdetail}
In confronting models with high scale physics inputs like 
SUGRA and string models with experiments, one must evolve the SUGRA and
string parameters including soft breaking parameters from a high scale down to the
electroweak region. This evolution leads in a natural way breaking of the 
$SU(2)_L\times U(1)_Y$ electroweak symmetry. This is the radiative 
electroweak symmetry
breaking (REWSB) scenario.  The resulting sparticle spectrum after REWSB 
carries the low energy signatures of the models. 
  Thus it is important to examine
the implications of REWSB in some detail. It turns 
out [\refcite{klcucpnprd1997}] that 
there are two regions of REWSB which are geometrically distinct and have 
widely different phenomenological implications. One of these is the 
ellipsoidal branch (EB)
and the other is the hyperbolic branch (HB). In the following we discuss 
the origin of these branches.

In the analysis of REWSB one must take into account loop corrections in 
the effective 
potential as these corrections can become important in certain regions 
of the parameter
space of models. Detailed numerical analyses, however, show that for the case
$\tan\beta<5$ the 
loop correction to the effective potential and well as the loop corrections to REWSB are small.  Now typically the REWSB  conditions arising from the minimization of the effective potential
give a constraint which is quadratic in the soft parameters $m_0$, $m_{1/2}$, $A_0$ and $\mu$.
Interestingly, for $\tan\beta<5$ the REWSB constraint for fixed $A_0$ 
and fixed $\mu$ gives an ellipse in the $(m_0-m_{1/2})$ plane. 
With $\mu$ having a lower limit, the implication of the constraint then
is that for fixed $A_0$ and fixed $\mu$ there are upper limits on $m_0$ and $m_{1/2}$. 

For $\tan\beta>5$ the loop correction to the effective potential and to REWSB is not small
and cannot be neglected.  Fixing $A_0$ and $\mu$ one finds that in this case $m_0$ and
$m_{1/2}$ lie on the boundary of a hyperbola.  
The qualitative difference between 
$\tan\beta<5$  and $\tan\beta>5$ cases can be understood as follows:
 For 
 $\tan\beta$ $>5$ , only $m_{H_2}^2$ rather than $m_{H_1}^2$ is important and 
the loop correction to the $\mu$ parameter, i.e., $\Delta \mu^2$ can be quite large at $M_Z$.  
It is seen that while $\mu^2$ and $\Delta \mu^2$ each have large dependence on the scale at
which is the potential is minimized,  their  sum $\mu_{tot}^2=\mu^2 +\Delta \mu^2$ remains 
reasonably flat with scale. It is then useful to  carry out a minimization of the one-loop 
effective potential at a scale where $\Delta \mu^2$ becomes negligible relative to $\mu^2$ 
({\em ie.} $\mu_{tot}^2 \sim \mu^2$). It  turns out that the specific point where the loop
correction turns out to be negligible lies  essentially midway between the largest and the 
smallest sparticle mass of the model. Further, this  scale is not far from the geometric  
mean of the  stop masses, i.e.,  $\sqrt{m_{{\tilde t}_1} m_{{\tilde t}_2}} $ for most of 
the parameter space of models. A minimization of the effective potential at the scale where
the one loop corrections are relatively small gives an insight into the REWSB constraint.
Thus at the above scale one finds that either the co-efficient of $m_0^2$ or the 
co-efficient of $m_{1/2}^2$ 
in the quadratic constraints  relating the soft parameters turns negative,  
thus changing an ellipse
into a hyperbola for fixed $\mu$ and $A_0$.  Here, both $m_{1/2}$ and $m_0$ may 
extend to  tens of TeV for  $\mu$ fixed. 
 The upper limit of $m_0$ for a given $m_{1/2}$ is then 
governed by the boundary defined by the lower limit of $m_{{\tilde \chi}_1^\pm}$ or 
REWSB.  This boundary  is quite sensitive to the top quark mass $m_t$. 
Another interesting insight is gotten if one interprets  $\mu^2/M_Z^2$ 
as a simple measure of fine-tuning.  In this case one finds that even a 
small fine tuning
can allow for large $m_0$ and $m_{1/2}$ on the hyperbolic branch.

\begin{figure}[hbt]
%\epsfxsize=10cm   %width of figure - will enlarge/reduce the figures
%\epsfbox{t30m0mh.eps}
%\figurebox{2cm}{3cm}{} %to have a box alone
%\centerline{\epsfxsize=4.1in\epsfbox{t30m0mh.eps}}
%\centerline{\epsfxsize=2.5in\epsfbox{t30m0mh.eps}}
%\centerline{\epsfxsize=2.5in\epsfbox{tt.ps}}
\begin{minipage}[b]{0.45\textwidth}
\epsfxsize=2.0in\epsfbox{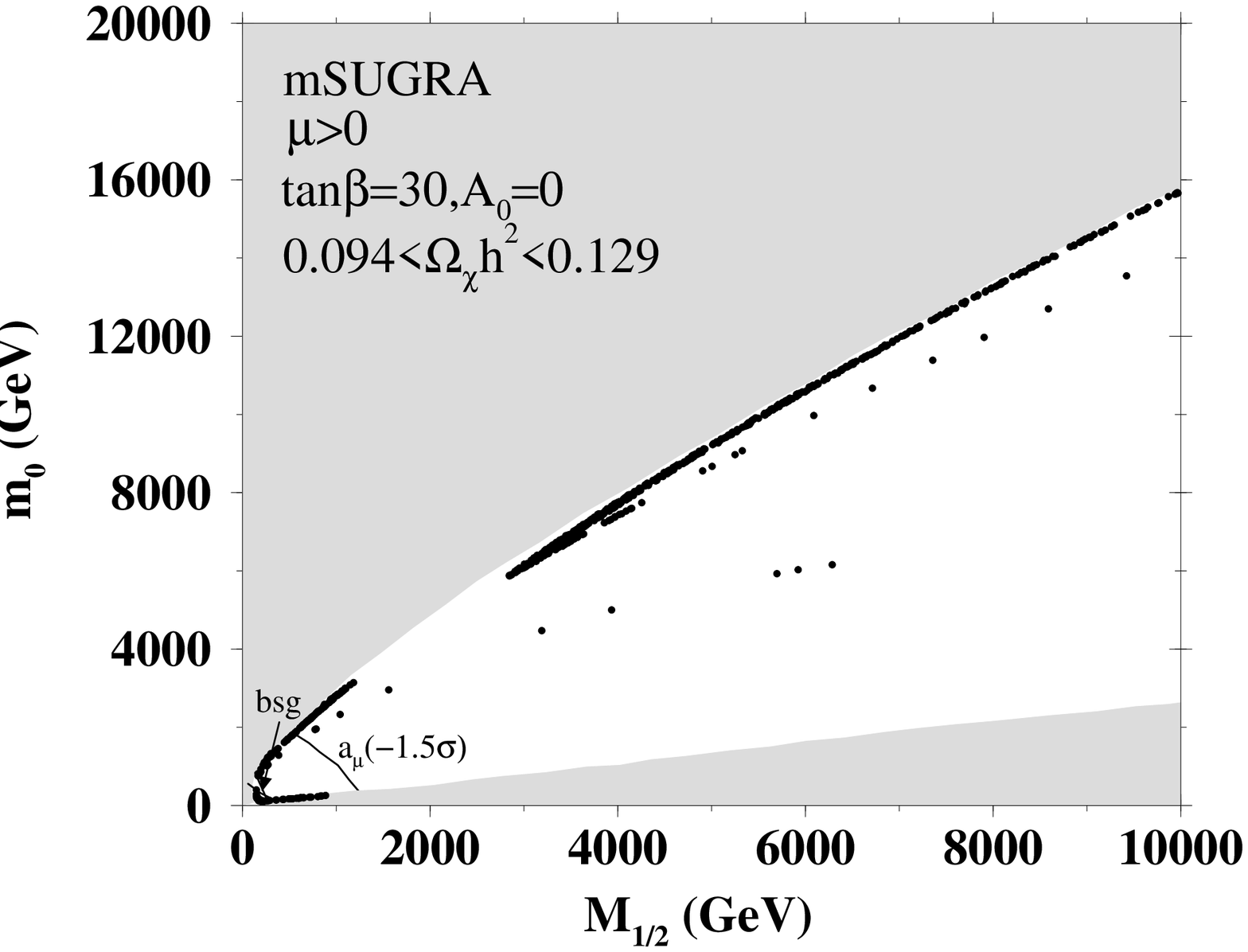}
\end{minipage}
\hspace*{0.1in}
\begin{minipage}[b]{0.45\textwidth}
\epsfxsize=2.0in\epsfbox{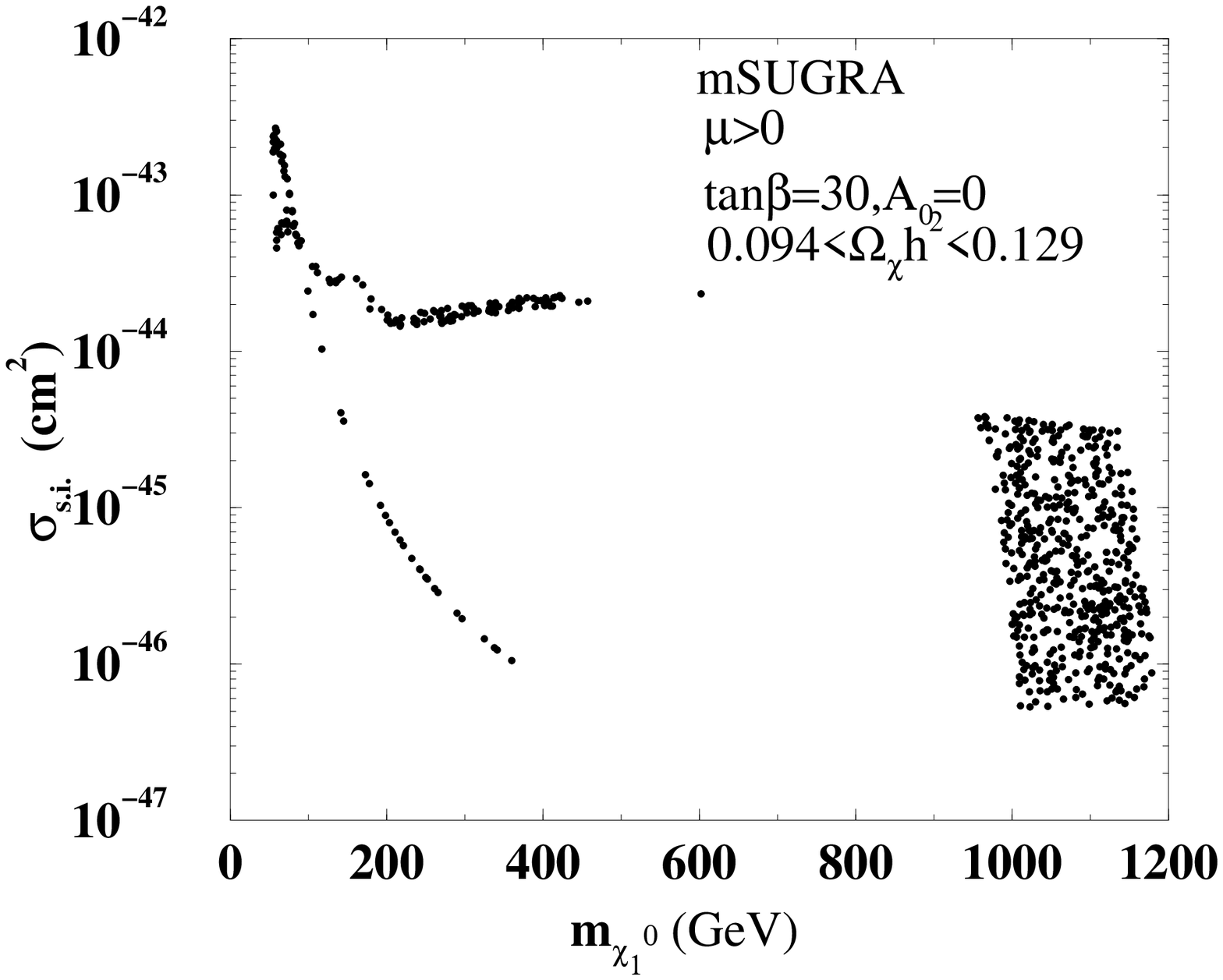}
\end{minipage}
\caption[]{Dark matter in the multi-TeV Hyperbolic branch/Focus Point 
scenarios (From Ref.\refcite{ucacpnprd2003}).  The WMAP satisfied 
relic density regions are shown in black.  The WMAP satisfied 
smaller $m_{1/2}$ region ($<1.5$~TeV) 
along the REWSB boundary (before the white region) 
falls in the focus point zone whereas the larger $m_{1/2}$ region ($>3$~TeV) 
along the same boundary falls in the inversion zone of the hyperbolic branch.     
}
\label{HBFPmSUG}
\end{figure}

The  hyperbolic branch was further analyzed in the context of dark matter in 
Ref.\refcite{ucacpnprd2003}. It is found useful here
to examine the composition of the LSP. One finds that
the gaugino- Higgsino content of the LSP  depends strongly on the region of the parameter 
space one is in. We consider the following distinct regions:
  (i) bino dominated scenario which typically arises in  mSUGRA with $m_0$ and $m_{1/2}$ 
 away from the low $\mu$ hyperbolic boundary. (ii) mixed gaugino-higgsino region 
which arises on the hyperbolic boundary with $m_{1/2}< 1.5$ ${\rm TeV}$. This is the focus
point region\cite{FengFocus} where $\mu$ is small.    
(iii) higgsino dominated scenario in the inversion region of the hyperbolic boundary with 
$m_{1/2}> 2.5$ ${\rm TeV}$ or so. 
The hyperbolic region of very large $m_0$ and $m_{1/2}$ is explored in 
Fig.\ref{HBFPmSUG}.  This resulted in the {\it inversion region} 
where the gaugino masses satisfy $m_i>> |\mu|$, for $i=1,2,3$.  The masses of  
squarks, gluino, heavier neutralinos, heavier chargino and 
heavier Higgs bosons can be 
very large (even several tens of TeV) in this region and fall 
outside the reach of the Large Hadron Collider (LHC).  
The only light sparticles in the system aside from the light Higgs are the 
 states ${{\tilde \chi}_1^0}$, ${{\tilde \chi}_2^0}$ and ${{\tilde \chi}_1^\pm}$
 which are almost degenerate in mass. Because of this it turns out to be important
 to include lowest order perturbative corrections to these masses for a reliable
 prediction of phenomena involving these states. 
  
Within the inversion 
region one finds $m_{ {\tilde \chi}_1^0} < m_{ {\tilde \chi}_1^\pm} <
m_{ {\tilde \chi}_2^0}$ at the tree level.  While each of the above masses 
lie in the range of several hundred GeV to about 1 TeV, the mass differences 
 $\Delta M^\pm=m_{ {\tilde \chi}_1^\pm}-m_{ {\tilde \chi}_1^0}$ and 
$\Delta M^0=m_{ {\tilde \chi}_2^0}-m_{ {\tilde \chi}_1^0}$ 
are quite small ($O(10)$ GeV or smaller). As already pointed out, 
in the computation of 
these 
mass differences radiative corrections\cite{pierce,drees} to the neutralinos 
and lighter 
chargino should also be included as they may have significant effects.
However, the smallness of these
mass differences poses a challenge regarding the observation of such particles.
Although the observation of sparticles at colliders may be difficult 
on the inversion  region of the hyperbolic
branch, it may still be possible to observe dark matter 
in this region. This topic
will be discussed in further detail in Sec.\ref{subsec:darkfocus}.

%\section{Precision Data Constraints on SUGRA and strings}
\section{Precision Data Constraints}
\label{muonG}
In a number of processes the supersymmetric loop correction for physical 
observables 
may be comparable to the standard model loop correction.  For such observables 
experiments that can probe the standard model correction can also be 
instrumental in testing corrections to models involving supersymmetry.  
%given by SUGRA and string. 
An example of such an observable is the muon anomalous magnetic moment
$a_{\mu}$\footnote{
$a_\mu$ is defined by the operator 
$\frac{e}{2m_\mu} a_\mu \bar \mu \sigma_{\alpha \beta} \mu F^{\alpha \beta}$.}.
Here it 
 was pointed out quite a while ago\cite{oldSusyMuong} that the supersymmetric 
electroweak contribution to the 
anomalous magnetic moment of the muon ($a_\mu^{SUSY}$) 
may be as large or larger than the 
standard model electroweak 
contribution\footnote{It turns out that the effect of extra compact 
dimensions on $a_{\mu}$ 
does not provide a big background to the supersymmetric correction and can be ignored 
in extracting the supersymmetric signal from data\cite{nyg2}.}.
Further, it was seen that the sign of $a_{\mu}^{SUSY}$ may 
be directly correlated with the sign of the 
$\mu$ parameter.  Thus precision experiments designed to 
test the standard model electroweak
correction would also test the supersymmetric correction, determine its sign and hence the
sign of the $\mu$ parameter and also constrain the parameter space of mSUGRA, extended SUGRA 
and other competing models of soft breaking. 
The one loop supersymmetric correction to $a_{\mu}$ arises from the chargino and the neutralino
exchanges. A detailed analysis of the supersymmetric correction within mSUGRA and other
scenarios was carried out in Ref.\refcite{ucpnprd1996} (see also 
Refs.\refcite{Lopez:1993vi},\refcite{Moroi:1995yh}) and the effect of CP phases was taken
into account in later works\cite{ing2}. 
These analyses led to several important observations. Within the mSUGRA it was 
shown that the lighter chargino-sneutrino 
(${\tilde \chi}_1^\pm-{\tilde \nu}_\mu$) loop 
is the dominating one relative to 
the heavier chargino-sneutrino (${\tilde \chi}_2^\pm-{\tilde \nu}_\mu$) loop and 
all the neutralino-smuon (${\tilde \chi}_i^0-{\tilde \mu}$) loops. 
The dominance of the light chargino exchange results in a strong $\tan\beta$ dependence 
of $a_\mu^{SUSY}$. It was further shown that 
$sign(a_\mu^{SUSY})=sign(\mu)$ in mSUGRA\footnote{The generalized  
result is $sign(a_\mu^{SUSY})=sign({\tilde m}_2 \mu)$.  We follow the 
usual sign conventions for $\mu$ and $A_0$ as in Ref.\refcite{signconvention}.}.
This particular relationship between the sign of $a^{SUSY}$ and the sign of $\mu$ is 
useful in the analysis of experimental data\cite{Brown:2001mg,ucpnprl2001}\footnote{See Ref.\refcite{2001muonG} for additional works.}.
 Additionally, in the theoretical predictions of $a_{\mu}^{SUSY}$  the 
  $b\rightarrow s+\gamma$ constraint also plays an important role.
The muon $g-2$ data can also be useful to constrain other SUSY breaking scenarios
like the minimal Anomaly Mediated Supersymmetry Breaking (mAMSB) 
model\cite{ucdkgsrprd2000}.

\begin{figure}[hbt]
%\epsfxsize=10cm   %width of figure - will enlarge/reduce the figures
%\epsfbox{t30m0mh.eps}
%\figurebox{2cm}{3cm}{} %to have a box alone
%\centerline{\epsfxsize=4.1in\epsfbox{t30m0mh.eps}}
%\centerline{\epsfxsize=2.5in\epsfbox{t30m0mh.eps}}
%\centerline{\epsfxsize=2.5in\epsfbox{tt.ps}}
\begin{minipage}[b]{0.45\textwidth}
\epsfxsize=2.0in\epsfbox{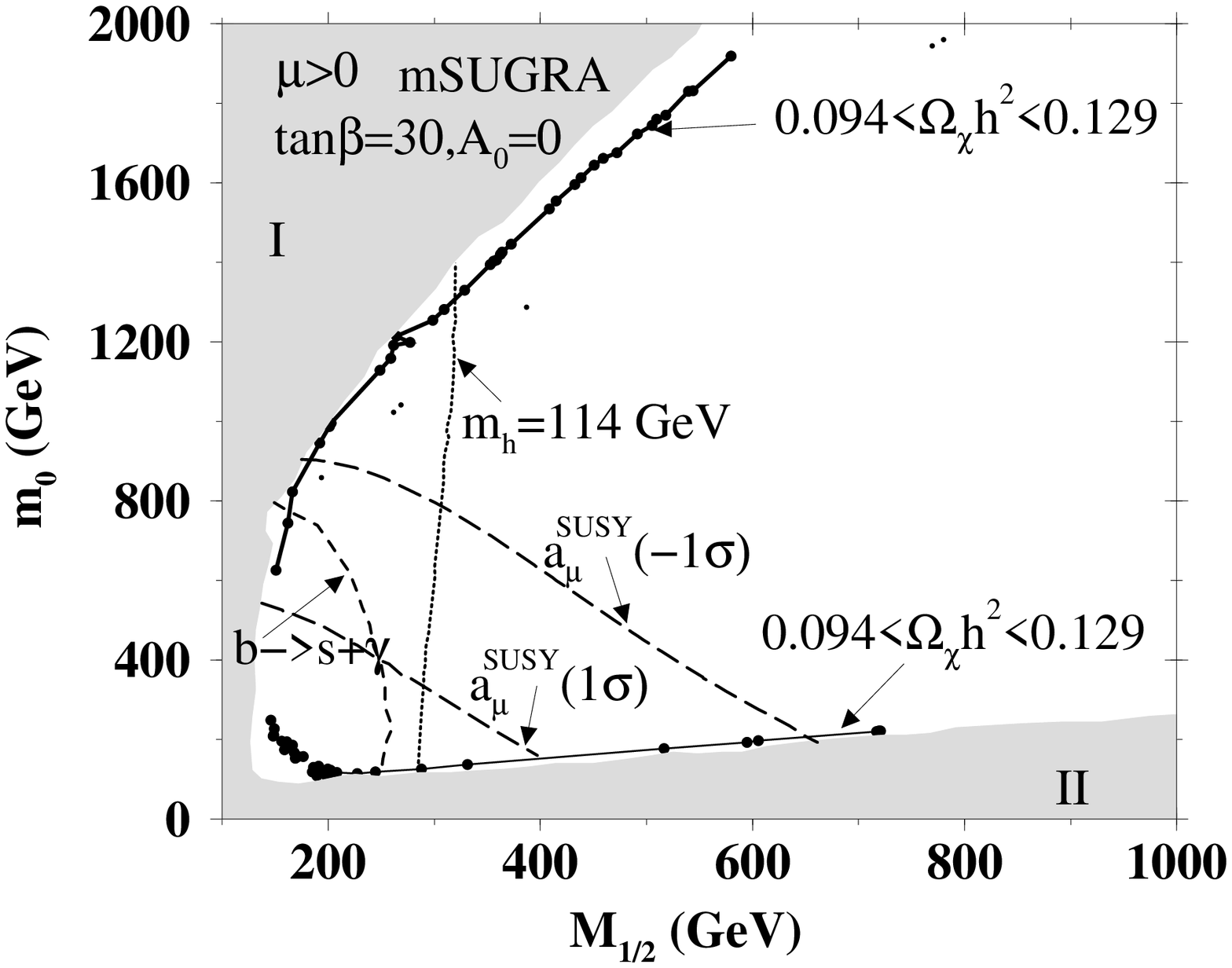}
\end{minipage}
\hspace*{0.1in}
\begin{minipage}[b]{0.45\textwidth}
\epsfxsize=2.0in\epsfbox{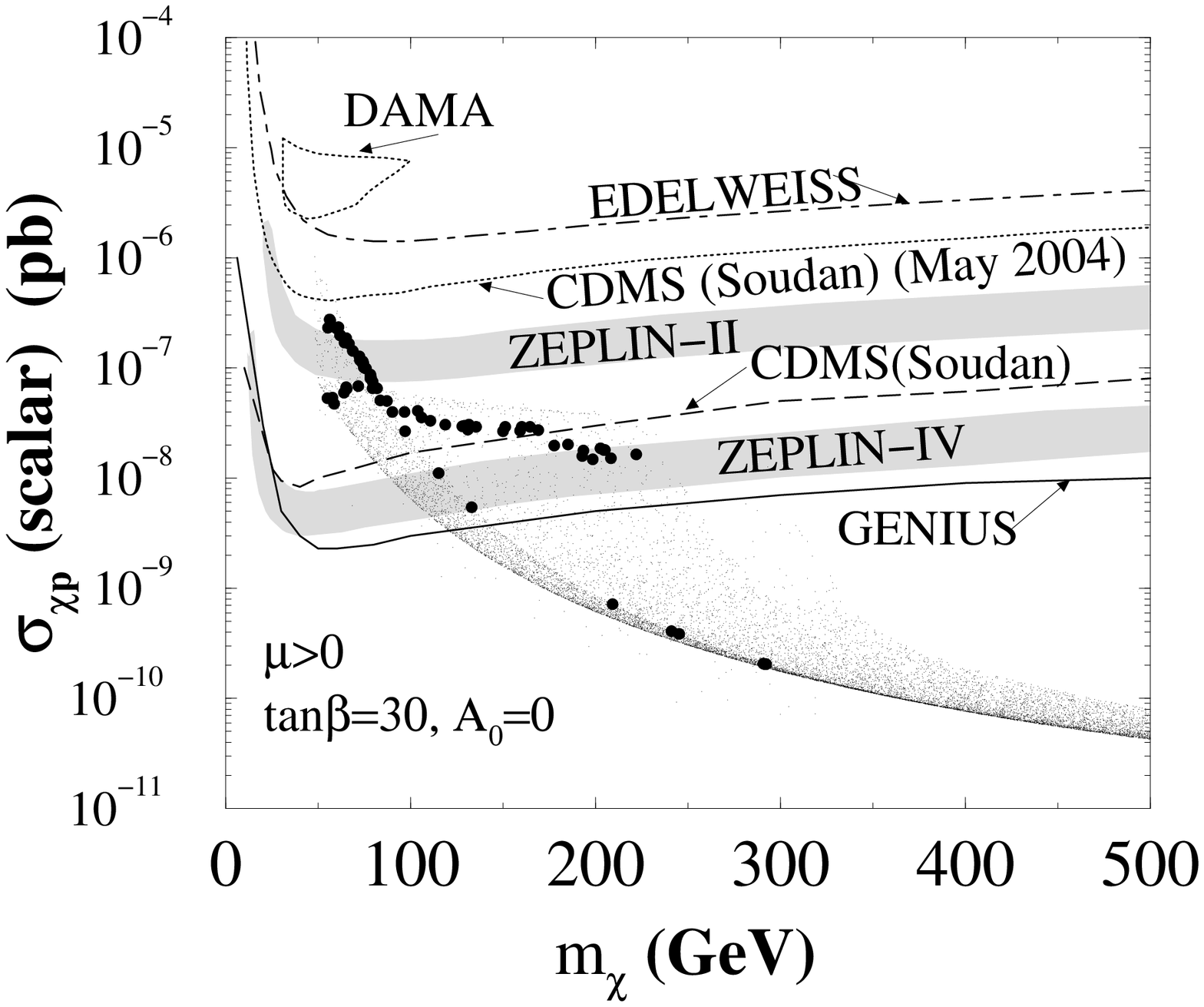}
\end{minipage}
\caption[]{Present constraints on mSUGRA $m_{1/2}-m_0$ plane 
and spin-independent ${\tilde \chi}_1^0-p$ cross section vs $m_{{\tilde \chi}_1^0}$ 
for $\tan\beta=30$. The black regions lines/circles are the valid 
relic density region from WMAP.  Contours are shown for the limits from  
$a_\mu^{SUSY}$ (via $\Delta a_\mu (e^+e^-)$ ) and 
from $Br(b\rightarrow s+\gamma)$ (left region disallowed) besides 
the LEP Higgs bound of $\sim$ 114 GeV\cite{lephiggs}. 
The present 
and future dark matter detection limits are also shown. The top gray 
region in the $m_{1/2}-m_0$ plane is discarded by the REWSB constraint and 
the bottom gray zone is the stau-LSP region.}
\label{presmSUG}
\end{figure}

Over the past few years the experimental 
accuracy of $g_{\mu}-2$ has increased very significantly. However, an extraction of new physics
signal from the data requires a comparably accurate estimate of the hadronic correction which
contributes to the standard model prediction. The computation of the hadronic correction has quite
an interesting history and currently it is still the most ambiguous part of the the standard model
prediction.  The hadronic correction to the standard model prediction has many components.  
It consists
of $O(\alpha^2)$ and $O(\alpha^3)$ hadronic vacuum polarization corrections, 
and corrections arising
from the light by light contribution. The light by light correction has undergone a dramatic shift
since it was shown that a switch of sign was needed\cite{knecht,hkrevised} 
over the previous evaluations. 
The other large source of error in the hadronic contribution comes from the $O(\alpha^2)$ vacuum
polarization contribution.  
There are two independent recent 
numerical evaluations of this contribution. One of these is from 
Hagiwara {\em et. al.}\cite{Hagiwara:2003da} 
and the other from  Davier {\em et. al.}\cite{Davier:2003pw}. 
Hagiwara {\em et. al.} used the 
low energy data of $e^+e^- \rightarrow hadrons$ to compute the hadronic 
vacuum polarization contribution to $a_\mu^{SM}$ while Davier et.al. 
used $\tau$ decay data to compute the same. Using these two estimates of the 
vacuum polarization
contributions, and including all other standard model contributions, one finds two different 
estimates for the difference between experiment (see Ref. (\refcite{Deng:2004jz})) 
and the standard model contribution.
Assuming that the entire difference between experiment and the standard model arises 
from the supersymmetric
contribution one finds the following two estimates for $a_\mu^{SUSY}$:
(i) $a_\mu^{SUSY}\equiv \Delta a_\mu (e^+e^-)=(23.9 \pm 10.0)\times 10^{-10}$, and
(ii) $a_\mu^{SUSY}\equiv \Delta a_\mu (\tau~decay)=(7.6 \pm 9.0)\times 10^{-10}$.  
Clearly, a $2.4\sigma$ deviation from the SM result arises using the analysis of  
 Hagiwara {\em et. al.}, whereas there exists no discrepancy   between 
 experiment  and the standard model result for the evaluation of Davier {\em et. al.}.
 Here we adopt the result of case (i) which uses the  $e^+e^-$ data.  
In Fig.(\ref{presmSUG}) an updated $1\sigma$ contours of $a_\mu^{SUSY}$ 
is shown using the constraint of case (i).  The conclusion of this 
analysis is quite similar to what was obtained in Ref.\refcite{ucpnprl2001}. 
 
\section{$h_b-h_\tau$ Yukawa Unification,  
$a_\mu^{SUSY}$, $b \rightarrow s+ \gamma$ and Nonuniversalities}
\label{btauunification}
Typically $h_b-h_\tau$ Yukawa unification  requires a negative 
$\mu$\cite{baggerMuneg,deboerMuneg}.  
This is so because $h_b-h_\tau$ unification demands $m_b(M_Z)$ 
in a range where there is a  negative SUSY loop correction $\Delta_b$ 
to the b-quark mass\cite{mbcorrections}.  
\begin{equation}
m_b(M_Z)=\lambda_b(M_Z){v\over {\sqrt 2}} \cos\beta (1+\Delta_b)
\label{bsusycorr}
\end{equation}
The dominant supersymmetric
contribution to  $\Delta_b$ comes from the gluino and chargino exchanges so that
$$ \Delta_b \simeq  \Delta_b^{\tilde g} +\Delta_b^{{\tilde \chi}^{\pm}}$$ 
where $\Delta_b^{\tilde g} \propto \tan\beta \mu M_{\tilde g}$ and $\Delta_b^{{\tilde \chi}^\pm } \propto \tan\beta
\mu A_t $. Each of these contributions are proportional to $\mu$ and 
it turns out that a negative $\Delta_b$ arises for
negative $\mu$.  Thus  $\mu<0$ is preferred in mSUGRA  for $h_b-h_\tau$ unification.
 This is somewhat at odds with the 
BNL $g-2$ data 
which appears to indicate a  $\mu>0$ (see, however, Refs.\refcite{BaerYukUni,RabyYukUni,YamaguchiYukUni}).
However, it is fairly easy to reconcile these two results if we allow for 
 non-universality of the gaugino  masses.
  This comes about as follows: we can choose $\tilde m_2$ 
 to be negative and $m_{\tilde g}$ to be positive. This leads to a positive  
 $a_\mu^{SUSY}$ while  $\Delta_b$ is negative. A solution of this type was analyzed in   
Ref.\refcite{ucpnbtauprd2002}. Such a phenomenon arises naturally in specific groups representations
in $SU(5)$ and $SO(10)$ to which the gauginos may belong[\refcite{ucpnbtauprd2002}].

Non-universality of gaugino masses  may arise from a non-trivial 
gauge kinetic energy function\cite{corsetti2,andersonNonuniv,eent} 
$f_{\alpha \beta}$ where $f_{\alpha\beta}$ transform according to a symmetric
product of the adjoint representation of the gauge group. Thus in general $f_{\alpha \beta}$ 
has a non-trivial field content. 
For SU(5), one gets the symmetric product as the sum of the 
following representations: 
\begin{equation}
(24 \times 24)_{sym}=1+24+75+200  
\label{su5breakup}
\end{equation}
In general one has ${\tilde m}_i(M_G)=m_{1/2} \sum_r C_r n_i^r$, where $n_i^r$ 
for $i=1,2,3$ is the characteristic of the representation $r$ and $C_r$ is the  
relative weight. As an example, let us suppose that one retains only the 24 plet representation
on the right hand side of Eq.(\ref{su5breakup}). In this case one gets the following ratio
of gaugino masses at the GUT scale: ${\tilde m}_3(M_G):{\tilde m}_2(M_G):{\tilde m}_1(M_G)=2:-3:-1$.  
We see now that $\tilde m_2$ and $\tilde m_3$ have opposite signs.  Thus if the gauginos
transform as the 24 plet of $SU(5)$ then one can have both 
 $h_b-h_\tau$ unification as well as a positive $a_\mu^{SUSY}$  for either
  sign of $\mu$.  However, the FCNC constraint from $b \rightarrow s+ \gamma$ 
eliminates more parameter space for $\mu<0$ than for $\mu>0$.  Including
both signs of $\mu$, $h_b-h_{\tau}$ unification then occurs for the following range
of parameters: $\tan\beta=15-45$ for $\delta_{b\tau}<0.3$, where 
$\delta_{b \tau}=(|\lambda_b-\lambda_\tau|)/\lambda_\tau$.  
For $\delta_{b \tau}<0.05$, one has $\tan\beta \sim 30$.

Next let us consider the $SO(10)$ case.
One has the following result for the symmetric product of the 
adjoint representations: 
\begin{equation}
(45 \times 45)_{sym}=1+54+210+770 
\label{so10}
\end{equation}
There are many ways in which the $SO(10)$ gauge group can break down to the Standard
Model gauge group. We consider the following two patterns 
\begin{equation}
SO(10) \rightarrow SU(4) \times SU(2) \times SU(2) 
\rightarrow SU(3) \times SU(2) \times U(1)
\label{case1}
\end{equation}
and
\begin{equation}
SO(10) \rightarrow SU(2)\times SO(7) \rightarrow SU(3) \times SU(2) \times U(1)
\label{case2}
\end{equation}
As an illustration suppose that 
the symmetric product transforms like the 54 plet on
the right hand side of Eq.(\ref{so10}). In this case Eq.(\ref{case1}) gives the 
following ratios for the gaugino masses at the GUT scale\cite{chamoun}
\begin{equation}
{\tilde m}_3(M_G):{\tilde m}_2(M_G):{\tilde m}_1(M_G)=1:-3/2:-1
\label{case3}
\end{equation}
Next suppose we consider that the pattern of case Eq.(\ref{case2}). In this case we have the following
ratio of gaugino masses 
\begin{equation}
{\tilde m}_3(M_G):{\tilde m}_2(M_G):{\tilde m}_1(M_G)=1:-7/3:1
\end{equation}
We note that in each of the two cases above $\tilde m_2$ and $\tilde m_3$ have opposite
signs. Thus it is possible to reconcile $h_b-h_{\tau}$ unification, $a_{\mu}^{SUSY}>0$ and
the $b\rightarrow s+\gamma$ constraint for these. An analysis of 
  SUSY dark matter constraints and 
detection prospects for dark matter for these cases are 
analyzed in Ref.\refcite{ucacpntbtauprd2002}. A further discussion of this
 topic will be given in Sec.~\ref{subsec:darkyukawa}.

%\section{Relic density and detection rates in SUGRA and Strings}
\section{Relic Density and Detection Rates}
\label{cdmanalyses}
The current astrophysical data strongly indicates the existence of 
cold dark matter in the universe\cite{recentDMreview}.  
The Wilkinson Microwave Anisotropy Probe (WMAP) experiment 
measured the parameters of the standard cosmological 
model\cite{wmapresults}. These are: 
$\Omega_b=0.044 \pm 0.004$, $\Omega_m=0.27 \pm 0.04$ and 
$\Omega_\Lambda=0.73 \pm 0.04$.  Here, $\Omega_{b,m}=\rho_{b,m}/\rho_c$ where 
$\rho_{b,m}$ is the the baryon (matter) density and 
$\rho_c(=3H_0^2/(8\pi G_N))$ is the critical mass density 
required to close the universe. Here    
$H_0$ is the Hubble parameter and $h=H_0/(100 {\rm ~km/s/Mpc})$~\footnote{$1~pc=
3.2615~{\rm light~year}=3.0856\times 10^{18}$~cm} amounts to  
$h=0.71^{+0.04}_{-0.03}$.  
$\Omega_\Lambda$ comes from the 
dark energy contribution.  
The cold dark matter density 
is then given by $\Omega_{CDM}h^2 = 0.1126^{+0.008}_{-0.009}$, which at 
$2\sigma$ level gives the following limit. 
\begin{equation}
0.094 <\Omega_{CDM}h^2 <0.129
\label{wmapdata}
\end{equation}  
A specially attractive CDM candidate in R-parity conserved scenario of 
supersymmetric models is  the lightest neutralino 
(${\tilde \chi}_1^0$)~\cite{goldbergDM}. 
In supergravity models ${\tilde \chi}_1^0$ 
becomes the LSP for most of the region of the 
parameter space (see Refs.\refcite{recentSUSYDMreview,recentDMreview} 
for recent reviews) and one may consider 
$\Omega_{CDM}\equiv \Omega_{{\tilde \chi}_1^0}= \rho_{{\tilde \chi}_1^0}/\rho_c$. 
We should note here that the upper side of 
$\Omega_{CDM}h^2$ is a strong limit but the validity of the lower 
bound becomes weak if we accept other candidates of dark matter. 
At  high temperature of the early universe 
($T>>m_{{\tilde \chi}_1^0}$), ${\tilde \chi}_1^0$ was 
in thermal equilibrium with its decay products.  The ${\tilde \chi}_1^0 {\tilde 
\chi}_1^0$ decay products 
are fermion pairs ($f \bar f$), gauge boson pairs ($W^+W^-$ \& $ZZ$), Higgs 
boson pairs ($hh,HH,AA,hH,hA,HA,H^+H^-$) or gauge boson-Higgs boson pairs 
($Zh,ZH,ZA$~\&~$W^\pm W^\mp$) with decays occurring 
through $s,t$ and $u$ channel 
diagrams\cite{jungmanReview}.

In determining the relic density of neutralinos at the current time we follow
the standard procedure. Thus as the universe cools, the annihilation rate 
falls below the expansion rate of the universe and ${\tilde \chi}_1^0$ 
moves away from thermal equilibrium (freeze-out).   
$\Omega_{{\tilde \chi}_1^0}h^2$ can thus be computed 
for its present value by solving the Boltzmann equation for 
$n_{{\tilde \chi}_1^0}$, the number density of the LSP 
in a Friedmann-Robertson-Walker universe.  Computing neutralino 
relic density most importantly involves computing $<\sigma_{eff} v>$, where 
$\sigma_{eff}$ is the neutralino annihilation cross section (which 
involves many final states, hence computationally it becomes quite intricate) and   
$v$ is the relative velocity between two neutralinos annihilating each other. 
If there are species with mass close to the LSP mass, then one must include  
coannihilation processes\cite{coanniSet1,coanniStop,mizuta,adsDARK,Edsjo:1997bg,coanniSet2}.

 In MSSM the lightest neutralino, i.e., the LSP, is composed of the
bino, the wino (which are superpartners of U(1) and the $SU(2)_L$ gauge bosons) 
and two Higgsinos\cite{HaberKaneEtc}.  
\begin{equation}
{\tilde \chi}_1^0=N_{11}\tilde B + N_{12}{\tilde W}_3 + 
N_{13}{\tilde H}_1^0 +N_{14}{\tilde H}_2^0
\end{equation} 
Here the coefficients $N_{ij}$ are elements of the matrix that diagonalizes the 
neutralino mass matrix. The 
gaugino fraction of ${\tilde \chi}_1^0$ is defined by $F_g=|N_{11}|^2 + 
|N_{12}|^2$. A ${\tilde \chi}_1^0$ is gaugino like if 
$F_g$ is very close to 1($>$0.9) , higgsino like if $F_g<0.1$.  
Otherwise it is a  gaugino-higgsino mixed state.

In mSUGRA the composition of the lightest 
neutralino in various regions of the parameter space is 
as follows: (i) bulk region: In the bulk region
of the $(m_{1/2}-m_0)$ plane {\em ie.} the region 
between the REWSB boundary (on the larger $m_0$ side)  
and the ${\tilde \tau}_1$ turning LSP boundary (on the smaller $m_0$ limit)
${\tilde \chi}_1^0$ is gaugino-like or more specifically 
bino-like. (ii) Focus Point region: In the  
focus point (FP) region of the hyperbolic branch the LSP is in general a gaugino-higgsino mixed state.
(iii): Inversion  region (IR): In the  
inversion  region of the hyperbolic branch the neutralino is essentially in  a
purely higgsino state.  Typically in the 
bulk parameter space of mSUGRA the relic density is usually too 
large to satisfy Eq.(\ref{wmapdata}) 
except in the {\it bulk annihilation region} characterized 
by low $m_0$ and low $m_{1/2}$ values (see Fig.\ref{presmSUG}).  Coannihilations reduce 
the relic density appreciably in specific regions of the parameter 
space so as to satisfy Eq.(\ref{wmapdata}).  
Stau coannihilations\cite{coanniSet1} may reduce the relic density appreciably 
to satisfy the WMAP data and the WMAP allowed region
falls near the ${\tilde \tau}$-LSP boundary region with 
smaller $m_0$ values.  These coannihilations processes are of the type 
${\tilde \chi}_1^0 \tilde \ell_R^a 
\rightarrow \ell^a \gamma, \ell^a Z, \ell^a h$,
$\tilde \ell_R^a \tilde \ell_R^b \rightarrow \ell^a \ell^b$,
 and $\tilde \ell_R^a \tilde \ell_R^{b*} \rightarrow \ell^a\bar \ell^b,
\gamma \gamma, \gamma Z, ZZ, W^+W^-, hh$. Here $\tilde l$ is effectively 
a stau. The coannihilations for the HB/FP/IR regions are described in 
Sec.\ref{subsec:darkfocus}. The neutralino-stop coannihilations however occur 
for very limited values of $A_0$\cite{coanniStop}. Relic density is also 
satisfied for large $\tan\beta (>45)$ cases when 
$m_A\sim 2m_{{\tilde \chi}_1^0}$~\cite{Afunnel} 
which is associated with a large annihilation of type 
${\tilde \chi}_1^0 {\tilde \chi}_1^0 \rightarrow A \rightarrow 
f \bar f$. 
\subsection{Dark Matter in the Focus Point  and Inversion  
 Regions  of  the Hyperbolic Branch}
\label{subsec:darkfocus}
  A study of neutralino relic density with the WMAP data along with 
using other phenomenological constraints  for the regions (i)-(iii) 
above was made in Ref.\refcite{ucacpnprd2003} (other post-WMAP 
analyses may be seen in Ref.\refcite{postWMAPsusyDMother}).
%  In the smaller $m_0$ regions near the boundary of the discarded region of the 
%parameter space where stau ${\tilde \tau}_1$ becomes the LSP or turns tachyonic, 
%coannihilations of ${\tilde \chi}_1^0 $ with ${\tilde \tau}_1$ reduces 
%$\Omega_{{\tilde \chi}_1^0}h^2$ strongly towards satisfying 
%Eq.~(\ref{wmapdata}). 
Computation of $\Omega_{{\tilde \chi}_1^0}h^2$ in the 
focus point region\cite{FengFocus} and the inversion region\cite{ucacpnprd2003} 
of the hyperbolic 
branch particularly shows the existence of strong coannihilations of the LSP with 
lighter chargino ${\tilde \chi}_1^\pm$.  Some of the dominant coannihilation 
processes in these region are\cite{Edsjo:1997bg,mizuta}:
${\tilde \chi}_1^0 {\tilde \chi}_1^{+}, {\tilde \chi}_2^0 
{\tilde \chi}_1^{+}\rightarrow u_i\bar d_i, \bar e_i\nu_i, AW^+,ZW^+,
W^+h$; ${\tilde \chi}_1^{+} {\tilde \chi}_1^{-}, {\tilde \chi}_1^0 {\tilde \chi}_2^{0}\rightarrow u_i\bar u_i, d_i \bar d_i,
W^+W^-$.  Having the smallest mass difference between coannihilating sparticles 
the process ${\tilde \chi}_1^0 {\tilde \chi}_1^{+}$ indeed dominates among 
the above channels.  
As a result 
$\Omega_{{\tilde \chi}_1^0}h^2$ is reduced appreciably so that it satisfies 
Eq.~(\ref{wmapdata}) or coannihilations may even reduce it further (below the lower 
limit of Eq.~(\ref{wmapdata})) thus causing ${\tilde \chi}_1^0$ 
to be a sub-dominant component of dark matter. Such {\it sub-dominant} region 
falls between the FP and the inversion region near the REWSB boundary.  The analysis 
of Ref.\refcite{ucacpnprd2003} also included the constraint from $b \rightarrow s+ \gamma$ 
and $a_\mu^{SUSY}$.  $B_s^0 \rightarrow \mu^+  \mu^-$ limit was 
also included for possible large $\tan\beta$ implications.  The LSP mass limit 
satisfying the WMAP data which arise from the neutralino-stau coannihilation 
is seen to be as large as 500 GeV for all $\tan\beta$. 
The same limit when the inversion region of HB is considered goes to about 
1200 GeV.  However, we should keep in mind that 
even a moderate amount of supersymmetric contribution to $(g-2)_\mu$ can 
eliminate the entire inversion region of the hyperbolic branch.  

Also of interest is the spin-independent ($\sigma_{\tilde \chi_1^0p}^{SI}$) and spin-dependent 
($\sigma_{\tilde \chi_1^0p}^{SD}$) neutralino-proton 
scattering cross 
sections\cite{directdetection,sugradark,recentDMdetect,recentSUSYDMreview,jungmanReview}. 
Among these the former has more current 
interest from experiments like CDMS, EDELWEISS, ZEPLIN and 
GENIUS\cite{darkexperiments}.  The results show that the FP region  
is quite accessible in future experiments, but the inversion region will 
not be probed effectively.  Typically a heavily 
higgsino-dominated region can be probed via indirect 
detection of dark matter\cite{indirectDMdetect1,indirectDMdetect2} 
experiments like IceCubes and ANTARES\cite{indirectDetectExp}.  
It is expected that a 
narrow band within the inversion region may be 
probed similarly for indirect detection.

\subsection{Dark Matter and Yukawa unification}
% $(g-2)_\mu$ and $b \rightarrow s+ \gamma$
%with nonuniversalities}
\label{subsec:darkyukawa}
Yukawa unification imposes additional constraints and there exist only few works which 
have taken into account such a constraint in the analysis of dark 
matter\cite{gomezYukawaDark,ucacpntbtauprd2002,profumoYukawaDark}.
Specifically in  Ref.\refcite{ucacpntbtauprd2002} an analysis was carried out with  
non-universal gaugino masses.  
Within the SU(5) scenario with non-universal gaugino masses 
only the representation $r_{24}$ allows
$h_b-h_\tau$ unification and satisfies $a_\mu^{SUSY}$ as well as 
$b \rightarrow s+ \gamma$ constraints. The sparticle spectrum is quite light in this 
case. Thus for $\mu>0$ and $\delta_{b \tau}<0.3$ one finds  
$m_{{\tilde \chi}_1^0} < 65 {~\rm GeV}$, $\Omega_{{\tilde \chi}_1^0}h^2$  lies in the 
desired range, and $\sigma_{\tilde \chi_1^0p}^{SI}$ also lies in a range  accessible 
to future dark matter experiments.  For the SO(10) case where the gaugino mass matrix 
transforms like the 54 plet representation of $(45\times 45)_{\rm sym}$, the analysis
give results similar to that of the $SU(5)$ case. In this case one finds that for $\mu>0$ and 
$\delta_{b\tau},~\delta_{tb},~\delta_{t\tau}<0.3$, one has  
$m_{{\tilde \chi}_1^0} < 80 {~\rm GeV}$ and the analysis of neutralino  relic density and
the  spin-independent cross-section 
$\sigma_{\tilde \chi_1^0p}^{SI}$ are similar to the $SU(5)$ case discussed above.
Regarding the collider reaches, all the sparticles in above scenarios 
can be completely probed at the LHC.

\subsection{Dark Matter with Modular Invariant Soft Breaking}
\label{subsec:darkmodular}
We discuss now dark matter within the context of an effective low energy supergravity
theory which has an $SL(2,Z)$ modular invariance associated with a large radius- small
radius duality symmetry. For simplicity we will assume that the K\"ahler potential depends
on the dilaton field $S$ and the K\"ahler moduli fields $T_i$ (i=1,2,3). It is then easily
seen that soft breaking in such a model\cite{Nath:2002nb} arising from spontaneous breaking of supersymmetry
such as the one that arises via a hidden sector in supergravity theories is also 
modular invariant. Further, quite remarkably one finds that the constraints of radiative
electroweak symmetry breaking in such a model 
determines $\tan\beta$[\refcite{ucpnprd2004,Nath:2002nb}].  
The phenomenon that $\tan\beta$ is no longer 
a free parameter but rather is a determined quantity under the constraints of REWSB should
apply to a broader class of models, for example, to soft breaking in models based on intersecting 
D branes\cite{korsnath}.
An analysis of dark matter within this class of models was carried out and  the neutralino 
relic density computed\cite{ucpnprd2004}.  Quite remarkably it is found that the modular invariant theory
with $\tan\beta$ no longer a free parameter can satisfy the WMAP constraints. Further,
it is found that the case with $\mu>0$ where the WMAP constraints are satisfied leads to 
a dilaton dominated region while the case with $\mu<0$ where the WMAP constraints are 
satisfied leads to a moduli dominated region. The $b\rightarrow s+\gamma$ constraint further
restricts the parameter space of the model.  For the $\mu>0$ case one finds that the 
WMAP and the $b\rightarrow s+\gamma$ constraints lead to upper bounds on the sparticle
masses and quite remarkably these lie within the reach of the LHC. However, this does not
hold for the $\mu<0$ case where the limits are much higher and some may lie outside
the reach of the LHC. 
It is also interesting to analyze the possibilities for the direct detection of dark 
matter predicted within modular invariant supergravity theory. Here an 
 analysis of the spin-independent neutralino-proton cross section ($\sigma_{\tilde \chi_1^0p}^{SI}$) 
shows that these cross sections can be  successfully probed in the current and future 
dark matter detectors [\refcite{ucpnprd2004}]. The analysis 
of Ref.\refcite{ucpnprd2004} 
is indeed the first work to use the dual constraints of modular invariance and 
radiative breaking of the electroweak symmetry for a determination of $\tan\beta$ 
and utilizes such a determination for the analysis of   
sparticle spectra and dark matter. 
[For other phenomenological analyses of soft-breaking using modular invariance see
 Ref.\refcite{otherModular}].   

\section{Conclusion}
    In this paper we have discussed possible tests of SUGRA and string 
    models using precision data and
    data from dark matter experiment. We also discussed the constraints of 
    Yukawa unification. The analysis was done in a variety of supergravity based 
    models, which include mSUGRA, extended SUGRA with gaugino mass non-universalities,
    and models with modular invariant soft breaking in generic heterotic string 
    models.  
    In comparing theory with experiment one must ascertain to a high degree of accuracy
    the predictions of the standard model so that one may reliably determine the 
    deviation between experiment and the standard model result. 
    The case in point is the BNL experiment
    which has recently produced a value of $(g_{\mu}-2)$ with a significant improvement
    over the previous determinations. However, a determination of whether or not 
    a new physics effect exists depends on a prediction of the standard model result
    to a comparable level which in turns depends on the accuracy of the hadronic correction.
    A significant new physics signal, i.e.,  at the level of $2.4\sigma$ results if one
    adopts the estimates of Hagiwara {\em et. al.} for the hadronic correction.  If one identifies
    supersymmetry as the origin of this deviation, then one finds that it leads to upper
    limits on sparticle masses within reach of the LHC for the $\mu>0$ case. 
   A positive $\mu$ is also favored 
by the $Br(b \rightarrow s+\gamma)$ constraint in the sense that the limits allow 
a large region of parameter space  with light sparticle masses which are just what 
one needs to generate a large supersymmetric correction to $a_{\mu}$. 
One drawback of a positive $\mu$ is that it is difficult to achieve Yukawa unifications 
for this case at least if one assumes universality of gaugino masses at the grand 
unification scale. 
This motivates us  
to explore non-universal gaugino mass scenarios where Yukawa unifications 
may be obtained for $\mu>0$ along with a satisfaction of  
the  muon $g-2$ and the $Br(b \rightarrow s+\gamma)$ constraints. The sparticle spectrum
in such scenarios again turns out to be low lying and can be probed by the LHC.  
Further, current and future dark matter detectors can also probe fully the 
dark matter predicted by this scenario in the direct detection dark matter experiments.

SUGRA and string models under the constraint of radiative electroweak symmetry breaking 
have two important branches: the ellipsoidal branch  and the hyperboloidal branch.
The former is valid for 
small $\tan\beta$~$(<5)$, whereas the latter exists for larger $\tan\beta$. 
Again the hyperbolic branch has two distinct sectors: the focus point (FP) region
and the inversion region (IR).  The focus point 
region  corresponds to  relatively low  values of  $m_0$ and $m_{1/2}$. 
 Here, the lightest neutralino has a significant amount 
of the Higgsino component.  On the other hand, the inversion region of the hyperbolic 
branch has relatively larger $m_0$ and $m_{1/2}$ 
while $\mu$ is still small. In this region the lightest neutralino is almost purely 
higgsino and most of the sparticles are heavy  with masses in the several TeV 
region except for
${{\tilde \chi}_1^0}$, ${{\tilde \chi}_2^0}$ and ${{\tilde \chi}_1^\pm}$. 
Quite remarkably on the hyperbolic branch the 
 relic density constraint from WMAP are still satisfied for both the focus 
point and the inversion region. Further, even though a large part  of this 
region may lie outside the reach of the LHC a significant part would 
still be accessible to dark matter experiments of direct detection type. 

 Finally we have also discussed in this review the low energy signatures of the modular invariant 
 soft breaking within the heterotic 
string frameworks while using the radiative electroweak symmetry breaking 
constraint. The use of the latter constrains the models significantly by 
fixing $\tan\beta$.  Remarkably the neutralino relic density satisfies 
the WMAP cold dark matter limits in a significant region of the parameter space 
of such models. Further the analysis leads to upper limits of sparticle masses for $\mu>0$ 
under the combined constraints of WMAP and  $Br(b \rightarrow s+\gamma)$.
It is observed that the 
  $\mu>0$ case leads to the allowed parameter space being  dilaton dominated and almost all 
the sparticles are found to lie within the reach of the LHC. Further, it is seen that the 
 direct detection experiments via 
spin independent scattering ($\sigma_{\tilde \chi_1^0p}^{SI}$) will 
completely probe such models.
 
\section{Acknowledgments}
  I thank the organizers of NathFest/PASCOS'04 for giving me an 
opportunity to write this review as a tribute to Pran Nath.

\end{document}